\documentstyle[preprint,aps,prl]{revtex}
\begin{document}
\draft
\title{
Coil Formation in Multishell Carbon Nanotubes: Competition between 
Curvature Elasticity and Interlayer Adhesion
}
\author{Ou-Yang Zhong-can, and Zhao-Bin Su}
\address{
Institute of Theoretical Physics, Chinese Academy of Sciences,\\ 
P.O. Box 2735, Beijing 100080, China}
\author{Chui-Lin Wang}
\address{China Center of Advanced Science and Technology (World Laboratory),\\ 
P.O. Box 8730, Beijing 100080, China} 
\date{\today}
\maketitle
\begin{abstract}
To study the shape formation  process of carbon nanotubes, 
a string equation describing the possible existing
shapes of the axis-curve of  
multishell carbon tubes (MCTs) is obtained in the continuum limit by 
minimizing the shape energy, that is
the difference between the MCT energy and the energy of the carbonaceous 
mesophase (CM).
It is shown that there exists a threshold relation of the outmost and 
inmost radii, that gives a parameter regime in which a straight
MCT will be bent or twisted. 
Among the deformed shapes,  
the regular coiled MCTs are shown being one of the solutions
of the string equation.
In particular,
the optimal ratio of  pitch $p$ and radius $r_0$ for such a coil 
is found to be equal to $2\pi $, which is in good agreement with
recent observation of coil formation in MCTs by Zhang
{\it et al.} 
\end{abstract}
\pacs{61.48.+c, 68.65.+g, 68.70.+w}

\narrowtext

Since the discovery of straight and multishell carbon nanotubes (MCTs) in
arc discharges \cite{1}, many unique and novel properties have been 
predicted for
the tubes. Among them, an especially intriguing one is their structural
stability, the mechanical properties of MCTs are expected to be significantly
stiffer than any presently known materials \cite{2}. However, in the recent 
synthesis by the catalytic decomposition of gas such as acetylene, 
a significant fraction of the produced MCTs exhibits various curved 
shapes \cite{3}, of which the most striking shapes are helices,
{\it i. e.}, the regular coils.
It has been pointed out that the tubes can be twisted  and deformed by 
an abrupt release of energy and a singularity in the stress-strain
curve \cite{4}.
In Ref. \cite{3},  the regular coil
formation was explained by a periodic distribution of pentagon-heptagon-pair
dislocations (PHPDs), but why the PHPDs are distributed periodicly rather
randomly still remains as an open question. Dunlap pointed out that the 
regularity could be caused by defect-defect interactions and constraints
on the optimal heptagon-pentagon nanotube bend \cite{4.5}.
In Ref. \cite{5}, the
bent carbon tubes were simulated by classical molecular dynamics based upon 
the three-body Tersoff-Brenner interatomic potential, 
but it is rather difficult to extract the essential physics from such a
numerical approach.

Therefore, the general questions can be posed as follows,
what is the mechanism of the curved deformation for the MCTs, and can we
derive the deformed shape.

In this Letter, we 
analytically obtain the general equilibrium-shape equation of the 
axis-curve of the MCT in the continuum limit
by taking account of competition among
the curvature elasticity, the adhesion of the interlayer van der Waals
bonding and the tension of the outer and inner surfaces of
a MCT. The sum of these three energies can be understood as the shape 
formation energy (see bellow for details).
We find that the variation of the shape formation energy yields
an equation of rigid string, which
has been studied extensively in differential geometry \cite{6}. Straight 
line and regular coil are two exact solutions of the equation. We show that 
under certain
geometric conditions the shape formation energy of a straight MCT
could become negative, in other words, the straight MCT 
becomes unstable in a quench-like formation process, and as a result,  
bent or 
twisted MCTs will be formed spontaneously to keep the equilibrium condition 
($i.~e.$ the zero shape formation energy).
Taking into consideration of the equilibrium condition in the 
quench-like cooling processes, the above argument provides an insight 
for the mechanism of the curved deformation for the MCTs.
The optimal ratio of the pitches and radii of the regular coils formed in 
such processes is shown impressively to be equal to $2\pi $. Our result 
is in good agreement with recent 
observation of the coil formation in MCTs by Zhang {\it et al.} \cite{3}. 

Generally, when the hydrocarbons are
thermally decomposited \cite{3}, the
carbon molecules are condensed mostly to form an isotropic smectic-like
crystal, {\it i. e.} the carbonaceous mesophase (CM), while the 
remained space is 
filled by plate-like molecules \cite{7}. Thus the MCT formation is quite 
similar to the tube
formation of a smectic-$A$ phase grown from isotropic phase in liquid 
crystal \cite{8}. As shown in Ref. \cite{8}, the shape formation energy is the 
additional energy of an MCT with respect to CM, which is a sum of the
following three terms, (i) the net difference of the volume free energy 
between MCT and CM, i.e. $F_V=-g_0V$ where $V$ is the volume of the MCT 
and $-g_0$ is
the adhesion energy density of the interlayer van der Waals bonding, (ii) the 
surface energy $F_A=\gamma (A_o + A_i)$ where $\gamma$ is the surface 
tension, 
$A_o$ and $A_i$ are the areas of the outmost and inmost surfaces, 
respectively, and (iii) the curvature elastic energy of the layers. 

We consider first the third term of the shape formation energy,
the curvature elastic energy.
MCTs can be treated as a set of curved graphite layers \cite{1}. 
For a single layer, the curvature elastic energy is an increment part of 
the in-layer covalent energy due to
the layer curvature.  Following Lenosky {\it et al.} \cite{9}, the
curvature elastic energy of a single layer curved graphite carbon has the 
form as, \begin{equation}
           E_b^{(s)}=\epsilon_1\sum_i (\sum_{<j>}{\bf u}_{ij})^2
   +\epsilon_2\sum_{<i,j>} (1-{\bf n}_i\cdot {\bf n}_j)+
  \epsilon_3\sum_{<i,j>} ({\bf n}_i\cdot
  {\bf u}_{ij})({\bf n}_j\cdot {\bf u}_{ji}),
                \label{eq1}
\end{equation}
where ${\bf u}_{ij}$ is a unit vector pointing from carbon atom $i$ to its
neighbour $j$ and ${\bf n}_i$ is a unit vector normal to the fullerene 
surface at atom $i$. The summation $\sum_{<j>}$ is taken over the three 
nearest neighbour $j$ atoms to atom $i$, and the sums of the last terms 
are taken over only the nearest neighbour atoms. The superscript $(s)$ 
emphasizes the energy $E_b$ is for a single layer. Our first task is to 
reduce Eq. (1) into a continuum form. 

A curved single-shell tube of radius $\rho$ without inclusion of its
two end-caps can be described  by
\begin{equation}
  {\bf Y}(s,\phi)={\bf r}(s)+\rho ({\bf N}(s)\cos \phi +{\bf b}(s)\sin\phi),
           \label{eq2}
\end{equation}
where $0<\phi<2\pi$, and $0< s < l$ is the arc-length parameter along the 
curved tube-axis whereas
curve ${\bf r}(s)$ is the vector representation for the curve of the tube
axis. $l$ is the total length of the tube. ${\bf N}(s)$ and
${\bf b}(s)$ are the unit normal and unit binormal vectors of ${\bf r}(s)$
respectively \cite{10}. Making use of the well-known Frenet formulas \cite{10},
\begin{equation}
{\bf t}_s = k(s){\bf N},\; \; \;  {\bf N}_s=-k(s){\bf t}-\tau (s){\bf b},  
\; \; \; {\bf b}_s=\tau (s){\bf N},
                \label{eq3}
\end{equation}
where ${\bf t}=d{\bf r}/ds$, ${\bf t}_s=d{\bf t}/ds$, ${\bf N}_s=d{\bf N}/ds$,
${\bf b}_s=d{\bf b}/ds$, $k(s)$ and $\tau(s)$ are the curvature and torsion of
${\bf r}(s)$ respectively. In ref. \cite{11}, we have derived 
the area element $dA=\rho (1-\rho k \cos \phi)d\phi ds$, the mean curvature
$H=(2\rho k\cos \phi -1)/2\rho (1-\rho k\cos \phi)$ and the Gaussian curvature
$K=-k\cos \phi / \rho (1-\rho k \cos \phi)$ 
for the tube surface of ${\bf Y}$. It is then easy to prove
$\oint KdA =0$. According to the Gauss-Bonnet theorem \cite{10}, this means 
the topology of the curved tube is the same as that of a straight tube. 
From the Euler's theorem in
topology, the surface can be perfectly embedded by a carbon network of 
six-member rings as in plane graphite layer. 
With the help of Frenet formula (3), we transform the vector 
functions in Eq. (1) into continuum limit by expanding them up to
the order of $O(a^2k^2)$,   
where $a=1.42\AA$ is the bond length in the graphite layer,
\begin{equation}
  {\bf u}_{i}(M)={\bf u}_{ij} =(1-\frac{a^2}{6}k^2(M)){\bf 
t}(M)+(\frac{a}{2}k(M)+
  \frac{a^2}{6}k_s(M)){\bf N}(M) - \frac{a^2}{6}k(M)\tau (M){\bf b}(M),
                \label{eq4}
\end{equation}
where $k_s=dk/ds$, 
$M=1, 2, 3$ denote three families of $sp^2$-bonded curves with one curve 
of each family acrossing from atom $i$ to its three neighbour atoms $j$
on the surface where the carbon atoms embedded, and have one to one 
correspondence to atoms $j$. 
Here we would like to emphasize that all the expressions $K(M)$, 
$t(M)$ ,$N(M)$, $\tau(M)$, and $b(M)$ are 
functions of the arc-length $s$, where $s = ia$. 
Specifying the tube surface described by Eq. (2), 
in which 
the $sp^2$-bonded curves can be considered approximately as geodesies 
for the curved surface ${\bf Y}$, we have  additionally 
${\bf N}(M)={\bf n}_i$, $k(M)=c_1 \cos ^2\theta (M)+c_2 \sin ^2 \theta (M)$,
$\tau (M)=(c_1-c_2)\sin \theta (M) \cos \theta (M)$ where $c_1$ and $c_2$ are
the two principal curvatures of the surface at atom $i$ location, $i.~ e.$,
$H=(c_1+c_2)/2$ and $K=c_1c_2$, and $\theta (M)$ are the angles between $c_1$
direction and ${\bf t}(M)$. Considering $\sum_{M=1}^3 \sin ^2\theta (M)=3/2$
and 
$\sum_{M=1}^3 \sin ^4 \theta (M)=9/8$, and substituting Eq. (4) into Eq. (1),
we obtain an important
formula for the curvature elastic energy of the tube \cite{11},
\begin{equation}
   E_b^{(s)}=\oint [\frac{1}{2}k_c(2H)^2+\overline{k} K]dA~,
             \label{eq5}
\end{equation}
where the bending elastic constant, 
\begin{equation}
   k_c=(1/32)(18\epsilon_1 +24\epsilon_2 +9\epsilon_3)(a^2/\sigma)
            \label{eq6}~,
\end{equation}
with $\sigma =\sqrt{3}a^2/4 =2.62\AA^2$ being the occupied area per atom, 
and the saddle-splay modulus is,
\begin{equation}
  \overline{k}=-(8\epsilon_2+2\epsilon_3)k_c/(6\epsilon_1+8\epsilon_2+
   3\epsilon_3)~.              \label{eq7}
\end{equation}
Formula (5) is actually a general expression of the elastic energy
which is valid also to fluid membranes \cite{12} and solid shells \cite{13}. 
If we substitute 
$(\epsilon_1, \epsilon_2, \epsilon_3)$ in Eqs. (6-7)  
by the values of $(0.96, 1.29, 0.05)$ eV respectively, which 
were calculated by Lenosky {\it et al.} 
using a local density approximation \cite{9}, we find $k_c=1.17$ eV and 
$k_c/\overline{k}=-1.56$. The obtained value of $k_c$ is in reasonable 
agreement with
the value of $1.02$ eV calculated  by Tersoff \cite{14} using an atomistic 
method for straight tubes, 
and is excellently close to the value of $1..2$ eV extracted 
from the measured phonon spectrum of graphite \cite{15}. The calculated ratio of 
$k_c/\overline{k}$ is also close to the result of 
$k_c/\overline{k}=-105.4/88=-1.2$ measured by Blakeslee {\it et al.\ } \cite{16}. 
Therefore, we have sufficient confidence on Eq. (5).
Moreover, since it has been averaged over three nearest neighbours for 
each site in expression Eq. (1) of Lenosky {\it et al.\ } \cite{9}, we can 
have only two invariants as $H^2 dA$ and $K dA$ in Eq. (5).
Therefore, within the same approximation, the free energy $F_b^{(s)}$ 
corresponding to Eq. (5) should be again a linear combination of these 
two invariants only.
Consequently, $F_b^{(s)}$ would have the same formal expression as Eq. (5) 
with coefficient $k_c$  and $\bar k$ being temperature dependent.

To extend the above result to a MCT, one has to integrate Eq. (5) from its
inmost radius $\rho_i$ to the outmost radius $\rho_o$. 
We may apply a similar treatment as that of \cite{17}
which is in fact devoted to
the curved smectic crystal multilayers, and has a layer structure
similar to graphites'. Replacing $k_{11}$ with $k_c/d$ in Eq. (3) of
Ref. \cite{17} and neglecting the
constant term associated with $\oint KdA$, we have
\begin{equation}
F_b=\sum F_b^{(s)}=(k_c/2d)\oint 2\sqrt{H^2-K}\ln (\frac{1-DH+D\sqrt{H^2-K}}{
1-DH-D\sqrt{H^2-K}})dA,     \label{eq8}
\end{equation}
where $d=3.4\AA$ is the space between two neighbour 
graphite layers and
$D=\rho_o -\rho_i$ is the thickness of the MCT. Here, the surface integral
is carried out over the inner surfaces. Using the above expressions for
$H$, $K$, and $dA$  and integrating from $\phi =0$ to $2\pi$, we
obtained the curvature elastic energy for the MCT as
\begin{equation}
F_b=(\pi k_c/d)\int \left[\ln (\frac{\rho_o}{\rho_i})+\ln (\frac{1+\sqrt{1-k^2
\rho_i^2}}{1+\sqrt{1-k^2\rho_o^2}})\right]ds.  
  \label{eq9}
\end{equation}

We now turn to consider the other two terms of the shape formation energy,
$F_V$ and $F_A$, both of which are weak binding energy,
$i.~ e.$ the adhesion energy between layers of an MCT.
Despite the fact that many of the structural properties of
plane graphites are well understood, the calculation of interlayer adhesion
energy for curved graphites is still an open question. The observation 
in Ref. \cite{1} reveals  that the interlayer
distance $d$ in MCTs remains to be the same as that in plane graphite, but 
the in-layer
lattice structures for each single-shell tube in one MCT may have different
helicity. In other words, the interlayer lattices are not in perfect
registry (referred to "incommensurate" or "mismatched" lattices). 
Therefore,
the attractive forces between layers in an MCT cannot be accounted for by 
conventional forces in plane graphites \cite{18}. However, the adhesion energy 
for mismatched
lattices is often smaller than that for commensurate surfaces. 
We use a mismatched parameter $\eta$ to take account for the 
mismatched effect between the interlayer lattices, $0 < \eta \leq 1$,
and $\eta = 1$ corresponds to the commensurate case. 
Since we are so far not aware of more detailed knowledges, as the lowest 
approximation, we take the following simple energy form,
\begin{equation}
F_V+F_A=-g_0\pi (\rho_0^2-\rho_i^2)\int ds+2\pi \gamma (\rho_0+\rho_i)\int ds,
      \label{eq10}
\end{equation}
where $-g_0=\eta \Delta E_c/d$, $\Delta E_c=-330 erg/cm^2=-2.04$ eV/nm$^2$
 is the interlayer cohesion energy of planar graphite obtained theoretically 
by Girifalco and Lad \cite{18},
and $\gamma$ is the
tension of the outmost and inmost surfaces 
of the MCT, which is equal to half of the energy needed to 
separate two unit-area 
surfaces, $i.~ e.$, $\gamma=-(1/2)\Delta E_c$.

Usually we have the MCTs with $\rho_o^2k^2 << 1$, then the expression  
$\ln [(1+\sqrt{1-k^2\rho_i^2})/
(1+\sqrt{1-k^2\rho_o^2})]$ can be 
approximated to  $(1/4)(\rho_o^2-\rho_i^2)k^2$, and
the shape formation energy of the MCT, Eqs. (9-10), can be 
subsequently simplified to,  
\begin{equation}
F=F_V+F_A+F_b=m\int ds +\alpha \int k^2ds~,
      \label{eq11}
\end{equation}
where $m=\pi (k_c/d)\ln (\rho_o /\rho_i)+2\pi \gamma (\rho_0+\rho_i)
-\pi g_0(\rho_o^2-\rho_i^2)$ and
$\alpha = (1/4)\pi (k_c/d)(\rho_o^2-\rho_i^2)$. Eq. (11) is nothing but a
string action \cite{19}. The variational equation 
$\delta F=0$ yields the
equilibrium-shape equations of the string \cite{6},  
\begin{equation}
  2k_{ss}+k^3-2k\tau ^2 -\frac{m}{\alpha} k=0,
     \label{eq12}
\end{equation}
\begin{equation}
   k^2\tau = \mbox{const.},
     \label{eq13}
\end{equation}
where $k_{ss}=d^2 k(s)/ds^2$.

Following what has been discussed for the derivation of curvature elastic
energy, in the derivation for these three terms of the shape formation
energy, we have expressed all the relevant quantities in terms of geometric
language. Therefore, due to the reason of geometric symmetry on the low
dimensional manifolds, the corresponding free energy should have the
same expression as those of the derived shape formation energy with the
coefficients becoming temperature dependent. We will keep such understand
in the above and hereafter discussions.

It is obvious that a straight line is always  a solution of 
the string equations (12) and (13), since its $k$ and $\tau $ are zero. 
The corresponding shape 
formation energy of a straight MCT is $F=ml$.
The shape formation energy is regarded as a free energy  and
the equilibrium threshold condition of
$F=0$ yields the criteria for the growth of a straight MCT as
\begin{equation}
m=\pi (k_c/d)\ln (\rho_o/\rho_i) +2\pi \gamma (\rho_o+\rho_i)-\pi g_0
(\rho_o^2-\rho_i^2)=0~.
      \label{eq14}
\end{equation}
This equation describes the geometric relation between $\rho_o$ and $\rho_i$ 
in terms of the physical parameters $k_c$, $\gamma$, and $g_0$ for a 
straight MCT. Both $\gamma$ and $g_0$
are also dependent on the formation temperatures and catalyst. So the 
detailed
data measured from the produced MCTs can reveal the properties of $\gamma$
and $g_0$ with the help of Eq. (14).
The formation procedure for MCTs, either a quick
growth in which the temperature can be regarded as constant 
or a sudden cooling, is actually a sort of quench-like process.
As long as the shape formation energy for the
straight MCTs, $i.~ e.$, deviates downwards from the threshold
condition in the formation procedure, becomes negative, the resultant
remnant part of energy will prevent the  straight MCT from keeping stable.
Then a shape deformation will be induced and it would lead to another solution
of the string equation with its shape formation again being equal to zero.
Therefore, by considering the threshold condition $F = 0$, any outward
growth by increasing $\rho_0$ will make the straight MCT undergo a shape
deformation as long as $g_0$ being kept constant. Furthermore, $g_0$ may
increase with temperature decreasing, following again Eq. (14),
a straight MCT grown may also be coiled under
the cooling process.
These features give a natural explanation for the
deformation of MCTs. 

Now we would prove that the observed MCT regular coils shown in Ref.\cite{3} are 
just the allowed 
solutions of equations (12) and (13).
Mathematically, the regular coils can be described  by vectors, 
\begin{equation}
{\bf r}(s)=(r_0\cos\omega s, r_0\sin\omega s, h\omega s),
     \label{eq15}
\end{equation}
where the coiled pitch $p = 2\pi h$, $r_0$ is the coil radius and 
$R \equiv \omega^{-1} = \sqrt{r_0^2+h^2}$. One can easily show from the 
Frenet formulas, Eq. (3), that  $k=\omega^2r_0$, $\tau
=-\omega^2h$, and the regular coil curves are the solutions of (12) and 
(13) if their $r_0$ and $h$ satisfy the following equation:
\begin{equation}
   r_0^2-2h^2-\frac{m}{\alpha}(r_0^2+h^2)^2=0.
        \label{eq16}
\end{equation}
Introducing  $h/R=\sin \theta$, and $r_0/R=\cos \theta$, and taking 
into account of Eq. (16),  we have
$R^2=(\alpha /m)(\cos^2\theta -2\sin^2\theta)$, and $k^2=r_0^2/R^4=
(m/\alpha)[1/(1-2\tan ^2\theta)]$. Therefore, the coil formation energy
can be derived from (11) as
\begin{equation}
F=ml [1+\frac{1}{1-2\tan^2\theta}]\; .
      \label{17}
\end{equation}
Since now we have the coil situation which is quite different from the 
straight MCT case, even for the negative value of $m$ in Eq. (17),  we may
treat the threshold condition $F = 0$ as
$\tan^2\theta =h^2/r_0^2=1$ or
\begin{equation}
\frac{p}{r_0}=2\pi\; .
    \label{eq18}
\end{equation}

We  compare the optimal ratio
given by Eq. (18) with the experimental results reported in Ref. \cite{3} and
find a good agreement. As shown in Fig.\ 1 of Ref. \cite{3}, 
there is a fraction (about $10\% $) of MCTs being regularly coiled with a 
variety of
radii $r_0$ and helix pitches $p$. By a direct evaluation from the figure, 
we do find $p/r_0=2\pi$ is well hold. 
A rough estimation from the coil shown in 
the inset of this figure gives $p\approx 600$ nm and $r_0\approx 100$ nm,
$i.~e.$ $p/r_0\approx 6\approx 2\pi$.
Another coil, shown in Fig.\ 2 of the same reference,  has its
$p\approx 700$ nm and
$r_0\approx 100$ nm which leads to $p/r_0 \approx 7$, again close
to the prediction of Eq. (18). Moreover, the results corresponding 
to the coil are shown in Fig.\ 3 of Ref.\ \cite{3}, 
 $\tan \theta=\frac{\pi}{2}-\phi_0\approx 2.2/2\approx 1.1$, are also in
good agreement with the present prediction of $\tan \theta 
=1$, where $\phi_0$ is defined within the context of Ref. \cite{3}. 

It is also interesting to study the value of $m$ for the MCT coil by
utilizing the data provided in Ref. \cite{3}, where the experimentally observed
values of $2\rho_i=3 \sim 7$ nm and $2\rho_o 
=15 \sim 20$ nm.
Making use of calculated values,  $k_c=1.17$ eV 
and $g_0=\eta \times 2.04\times 10^{-2}(eV/\AA^2)/d
= 6.03\eta$ eV/nm$^3$,
we can estimate $\pi g_0(\rho_o^2-\rho_i^2)= (1.0\eta \sim 1.7\eta$) $\times 
10^3$eV/nm, and $\pi (k_c/d)\ln (\rho_o/\rho_i)+2\pi 
\gamma(\rho_o+\rho_i)=68 \sim 101$ eV/nm.
Considering  expression (14), we find that under reasonable
approximation, $m$ in the practically formed MCT coils always take negative 
value.  Such a fact is notably consistent with the above proposed
explanation for the shape deformation mechanism of MCTs.

In summary, by deriving a string action type  expression for the 
formation energy of the MCTs as well as its equilibrium-shape equation, we 
have shown that there is a threshold condition for the formation of straight
MCTs, below that the straight MCTs become unstable and will undergo a 
shape deformation. In particular, we derive further an optimal formation
condition $p/r_0 = 2\pi$ for the regular coil solution which is
in good agreement with the recent experiment observations.

This work is partly supported by the National Natural Science Foundation of 
China.


\begin{references}
\bibitem{11} The detail will appear in a full paper by the present authors.
\end{references}
\end{document}